\documentclass[aip,pop,reprint]{revtex4-1}
\usepackage[pdftex]{graphicx}
\usepackage{graphicx}
\usepackage{amsmath}
\usepackage{amsfonts}
\usepackage{mathtools}
\usepackage{geometry}
\usepackage[caption=false]{subfig}
\usepackage[usenames,dvipsnames]{color}
\begin{document}

\title{Taylor-Couette Flow of Unmagnetized Plasma}

\author{C. Collins}
\affiliation{Department of Physics, University of Wisconsin, Madison WI 53706, USA}
\affiliation{Center for Magnetic Self Organization, University of Wisconsin, Madison WI 53706, USA}
\author{M. Clark}
\affiliation{Department of Physics, University of Wisconsin, Madison WI 53706, USA}
\author{C.M. Cooper}
\affiliation{Department of Physics, University of Wisconsin, Madison WI 53706, USA}
\affiliation{Center for Magnetic Self Organization, University of Wisconsin, Madison WI 53706, USA}
\author{K. Flanagan}
\affiliation{Department of Physics, University of Wisconsin, Madison WI 53706, USA}
\affiliation{Center for Magnetic Self Organization, University of Wisconsin, Madison WI 53706, USA}
\author{I.V. Khalzov}
\affiliation{Department of Physics, University of Wisconsin, Madison WI 53706, USA}
\affiliation{Center for Magnetic Self Organization, University of Wisconsin, Madison WI 53706, USA}
\author{M.D. Nornberg}
\affiliation{Department of Physics, University of Wisconsin, Madison WI 53706, USA}
\affiliation{Center for Magnetic Self Organization, University of Wisconsin, Madison WI 53706, USA}
\author{B. Seidlitz}
\affiliation{Department of Physics, University of Wisconsin, Madison WI 53706, USA}
\author{J. Wallace}
\affiliation{Department of Physics, University of Wisconsin, Madison WI 53706, USA}
\author{C.B. Forest}
\affiliation{Department of Physics, University of Wisconsin, Madison WI 53706, USA}
\affiliation{Center for Magnetic Self Organization, University of Wisconsin, Madison WI 53706, USA}

\date{\today}

\begin{abstract}

Differentially rotating flows of unmagnetized, highly conducting plasmas have been created in the Plasma Couette Experiment.  Previously, hot-cathodes have been used to control plasma rotation by a stirring technique [C. Collins \emph{et al.}, Phys. Rev. Lett. {\bf108}, 115001(2012)] on the outer cylindrical boundary---these plasmas were nearly rigid rotors, modified only by the presence of a neutral particle drag.  Experiments have now been extended to include stirring from an inner boundary,  allowing for  generalized  circular Couette flow and opening a path for both hydrodynamic and magnetohydrodynamic experiments, as well as fundamental studies of plasma viscosity. Plasma is confined in a cylindrical, axisymmetric, multicusp magnetic  field, with $T_e< 10$ eV, $T_i<1$ eV, and $n_e<10^{11}$ cm$^{-3}$. Azimuthal flows (up to 12 km/s, $M=V/c_s\sim 0.7$) are driven by edge ${\bf J \times B}$ torques in helium, neon, argon, and xenon plasmas, and the experiment has already achieved $Rm\sim 65$ and $Pm\sim 0.2 - 12$. We present measurements of a self-consistent, rotation-induced, species-dependent radial electric field, which acts together with pressure gradient to provide the centripetal acceleration for the ions. The maximum flow speeds scale with the Alfv\'{e}n critical ionization velocity,  which occurs in partially ionized plasma. A hydrodynamic stability analysis in the context of the experimental geometry and achievable parameters is also explored.

\end{abstract}


\maketitle

\section{Introduction}
Circular Couette flow (or Taylor-Couette flow, hereafter TCF) is one of the most fundamental systems in fluid dynamics \cite{taylor_1923, Chandrasekhar_1961, andereck_1986} and was first investigated over 130 years ago.  In TCF geometry, a viscous liquid flows between two coaxial, differentially rotating cylinders. The classic TCF profile is purely azimuthal, though experiments in finite height cylinders often suffer from secondary, non-azimuthal Ekman circulation due to end effects.  While seemingly simple, TCF exhibits a large number of distinct flow states (which depend on the rotation rates of the inner and outer cylinders), making it the ideal testbed for detailed experimental and theoretical studies of hydrodynamic instability thresholds, nonlinear behavior, and transition to turbulence \cite{Swinney_1978}. Recently, TCF of a conducting medium, such as liquid metal or plasma, has been pursued as a model for understanding astrophysical systems, and theoretical investigations using both ideal magnetohydrodynamics (MHD) and extended MHD in plasma \cite{ebrahimi2011_pop} have been examined. 

A TCF has recently been created using plasma, rather than liquids, in a laboratory device named the Plasma Couette Experiment (PCX) located at the University of Wisconsin-Madison. The goal of PCX is to create a sufficiently hot, steady-state, differentially rotating, weakly magnetized plasma to study basic hydrodynamic and magnetic flow-driven instabilities and dynamics specific to plasmas, such as two-fluid effects and neutral collisions. In PCX, toroidal plasma flow is controlled by electrostatic stirring assemblies at both the inner and outer boundaries. Flow profiles can be adjusted so that the angular velocity decreases with radius and the angular momentum increases with radius, and thus mimic the Keplerian-like flows of accretion disks (where $\Omega(r)=\sqrt{GM/r^3}$ ).

A TCF of plasma operates in a regime distinct from either liquid or liquid metal TCF. Specifically, the degree of ionization introduces the possibility of momentum loss through collisions with neutral atoms, and the resistive and viscous dissipation varies with the plasma temperature $T_e$ and density $n_e$. The relevant hydrodynamic Reynolds number characterizing the ratio of the momentum advection by the flow to the momentum diffusion, $Re=VL/\nu$, is governed by the flow velocity $V$, the characteristic size $L$, and Braginski viscosity; in an unmagnetized plasma, $\nu = 0.96 v_{T_i}^2 \tau_i$ where $v_{T_i}$ is the ion thermal speed and $\tau_i$ is the ion-ion collision time. Also of interest is the magnetic Reynolds number, $Rm=VL/\eta$, which describes the degree to which magnetic fields are frozen into a moving plasma with magnetic diffusivity $\eta$. In terms of plasma parameters,
\begin{eqnarray}
Re   &=&  0.74\; V_{\text{km/s}} L_ \text{m} n_{ \text{i,}10^{11}  \text{cm}^{-3}}Z^4\sqrt{\mu}\;T_{ \text{i,eV}}^{-5/2}\qquad\\
Rm  &=& 0.84 \; T_{e,eV}^{3/2} V_{km/s} L_m Z^{-1} 
\end{eqnarray}
The ratio of these quantities is the magnetic Prandtl number $Pm=Rm/Re=\nu/\eta$ and is of order unity in PCX plasmas,  a significant advantage over liquid metal experiments where $Pm=10^{-5}-10^{-6}$.
 
If rotation is driven exclusively at the inner boundary, TCF can become hydrodynamically unstable above a certain critical threshold (set by the viscosity) and consequently exhibits large-scale, secondary flows in the form of axisymmetric ``Taylor'' vortices, first studied in detail by G.I. Taylor in 1923 \cite{taylor_1923}. Exploring the onset threshold for hydrodynamic instability is important in PCX, since Taylor vortices govern the rate of momentum transport across the sheared flow of plasma. Envisioning the use of this device for experiments concerning other instabilities, such as the magnetorotational instability or generating a dynamo in TCF ~\cite{willis_2002}, requires a thorough understanding of the important dynamical effects and thresholds for hydrodynamic instability.

This article reports on a number of recent advancements to the experiment. Uniform, solid-body rotation spanning the entire height of the plasma is achieved by extending the rotation drive system to include all four cathodes at the outer boundary. Plasma parameters such as velocity, $T_e$, $n_e$, $V_{float}$ are measured with probe-based diagnostics, and a non-perturbative method of line ratio spectroscopy is also used to measure $T_e$. A newly constructed center post assembly provides stirring on the inner boundary, resulting in a controlled, differentially rotating flow which we use to explore the relevant dynamics of plasma TCF. Measurements of a self-consistent, rotation-induced radial electric field are also presented. Much effort was made to maximize the induced
flow, and apparent velocity limits have been observed for various gas species (He $\sim$ 12 km/s, Ne $\sim$ 4 km/s, Ar
$\sim$ 3.2 km/s, Xe $\sim$ 1.4 km/s), that scale with a critical ionization velocity limit reported to occur in partially
ionized plasmas. Finally, a hydrodynamic linear stability analysis is performed to estimate the critical velocity for the appearance of Taylor vortices at current experimental geometry and parameters.

\begin{figure}[t]
\begin{center}
\includegraphics[width=\columnwidth]{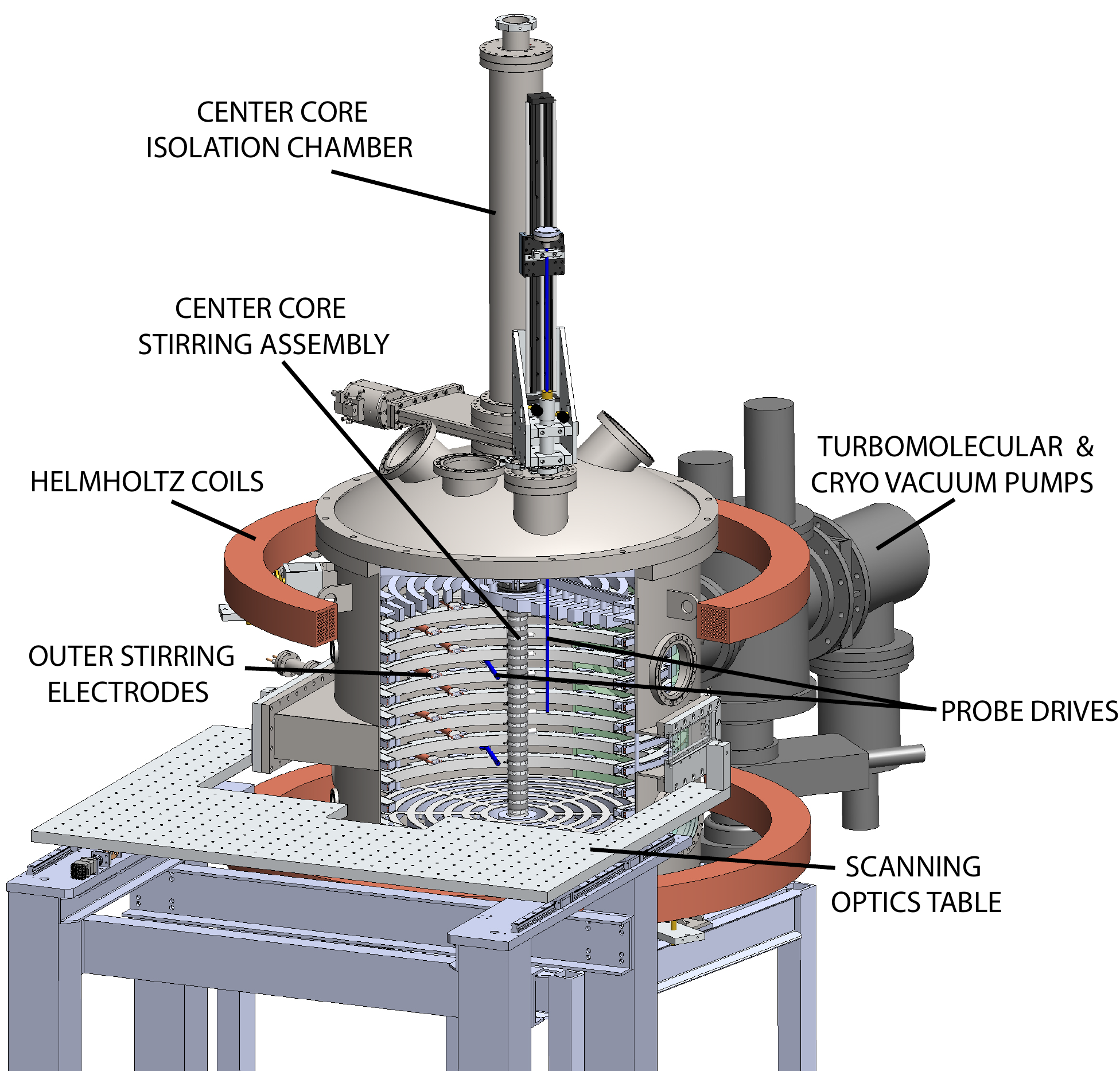} 
\caption{Major hardware components and cut-away view of the vacuum chamber.}
\label{fig:PCXschematic}
\end{center}
\end{figure}

\section{Description of Experiment}

\begin{figure}
\begin{center}
\includegraphics[width=.9\columnwidth]{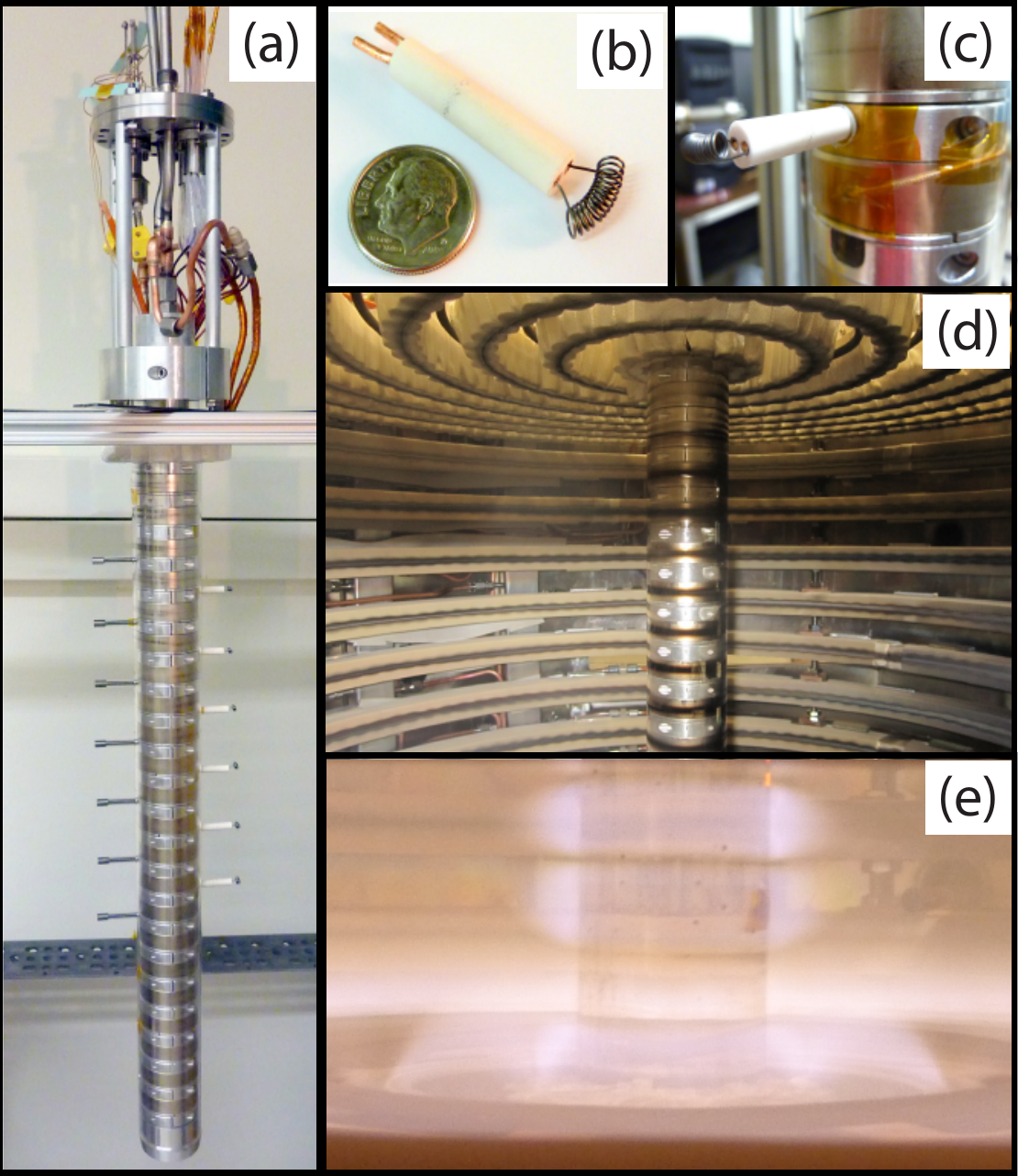} 
\caption{Newly installed center core assembly of magnets and stirring electrodes.}
\label{fig:PCXcollage}
\end{center}
\end{figure}

\begin{figure}[t]
\begin{center}
\includegraphics[width=.45\textwidth]{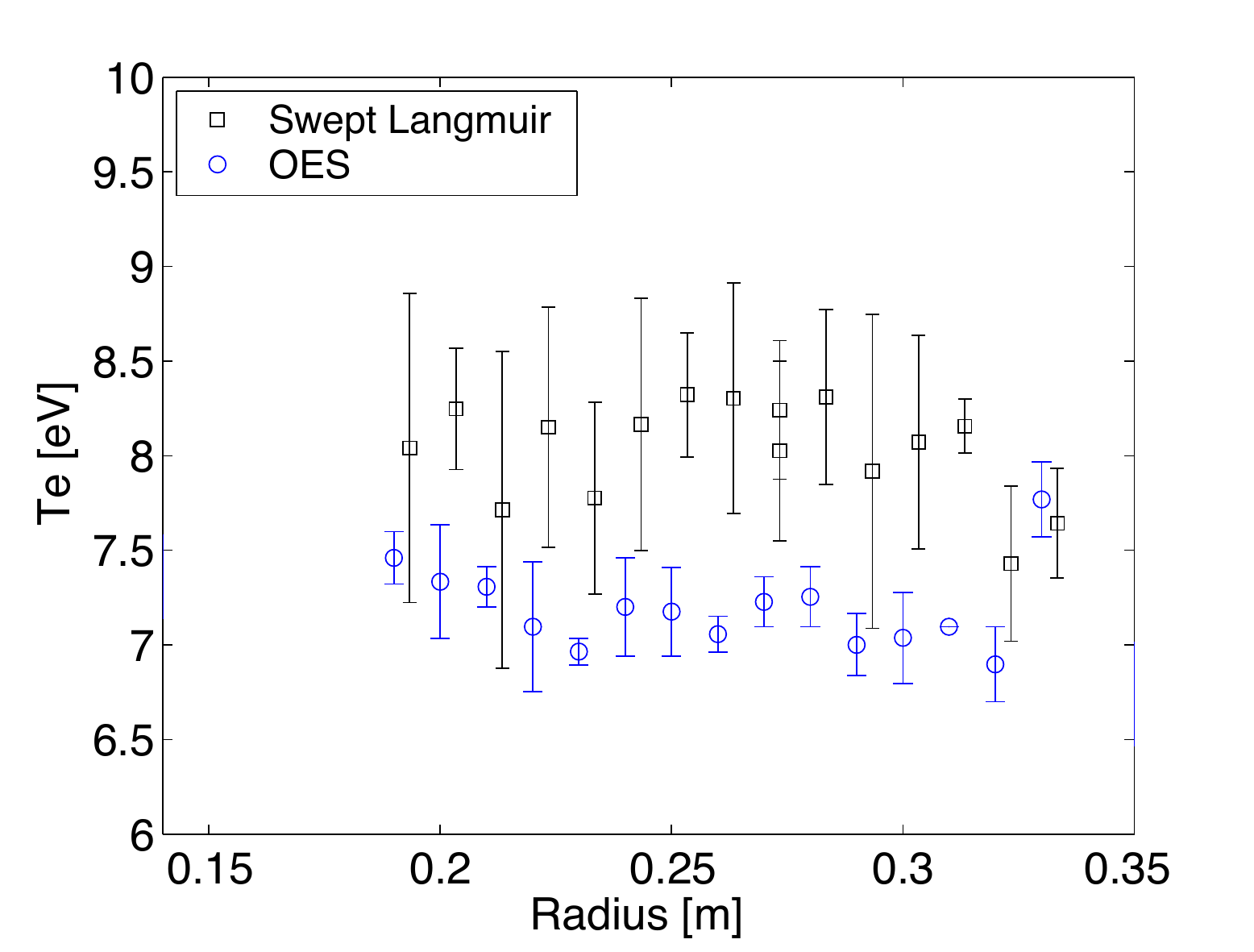}
\end{center}
\caption{Electron temperature vs. radius measured by the swept Langmuir probe and Optical Emission Spectroscopy suggest uniform temperature in the bulk, unmagnetized region ($0.1 < r < 0.3$ m).  Measurements were taken in a flowing argon plasma with 3 km/s solid body rotation (similar to Fig. \ref{fig:radialpotential}e). Error bars are standard error of time-averaged measurements. }
\label{fig:shot690te} 
\end{figure} 

In PCX, plasma is confined by a cylindrical ``bucket'' assembly of permanent magnets, arranged in rings of alternating polarity, to form an axisymmetric cusp magnetic field \cite{katz2012_rsi}. The field is localized to the boundaries, leaving a large, unmagnetized plasma in the bulk, with aspect ratio $\Gamma=height/(R_2-R_1) \lesssim 3$. Plasma is produced with 2.45 GHz microwave heating to reach $T_e<$ 10 eV and $n_e=10^{10}$-$10^{11}$ cm$^{-3}$. The waveguide launches mostly O-mode, implying a cutoff density limit at $n_e=7.4 \times 10^{10}$ cm$^{-3}$. Similar to other microwave powered multicusp plasmas \cite{Tuda2000}, the PCX plasma is observed to have two distinct states. In the overdense state, the plasma is produced by a surface wave discharge, exhibiting bright emission from the central, field-free region of the plasma, and cusp losses can clearly be seen.  When the plasma is underdense, bright rings of emission are observed, especially around the center core magnets near the ECH resonant zone (Fig.~\ref{fig:PCXcollage}e), indicating resonant electron heating and trapping. These emission rings are absent at locations where center post electrodes are installed due to increased loss area.

Electron temperature is routinely measured with single-tip swept or triple-tip Langmuir probes and has been confirmed using Optical Emission Spectroscopy (OES), a non-invasive diagnostic which uses emission line ratios to determine the electron temperature. The OES method\cite{boffard2010_psst} has previously been developed by collaborators on an inductively coupled argon plasma used in plasma processing with $1 <T_e < 6$ eV , but this is the first application to plasma with both higher electron temperature (up to 15 eV) and ionization fraction (30\%). The OES electron temperature measurements in PCX (Fig. \ref{fig:shot690te}) agree within 15$\%$ of Langmuir probe measurements \cite{Collins_thesis}.

\begin{figure*}
\begin{center}
\subfloat[]{\label{fig:zfil2or4a}\includegraphics[width=\columnwidth]{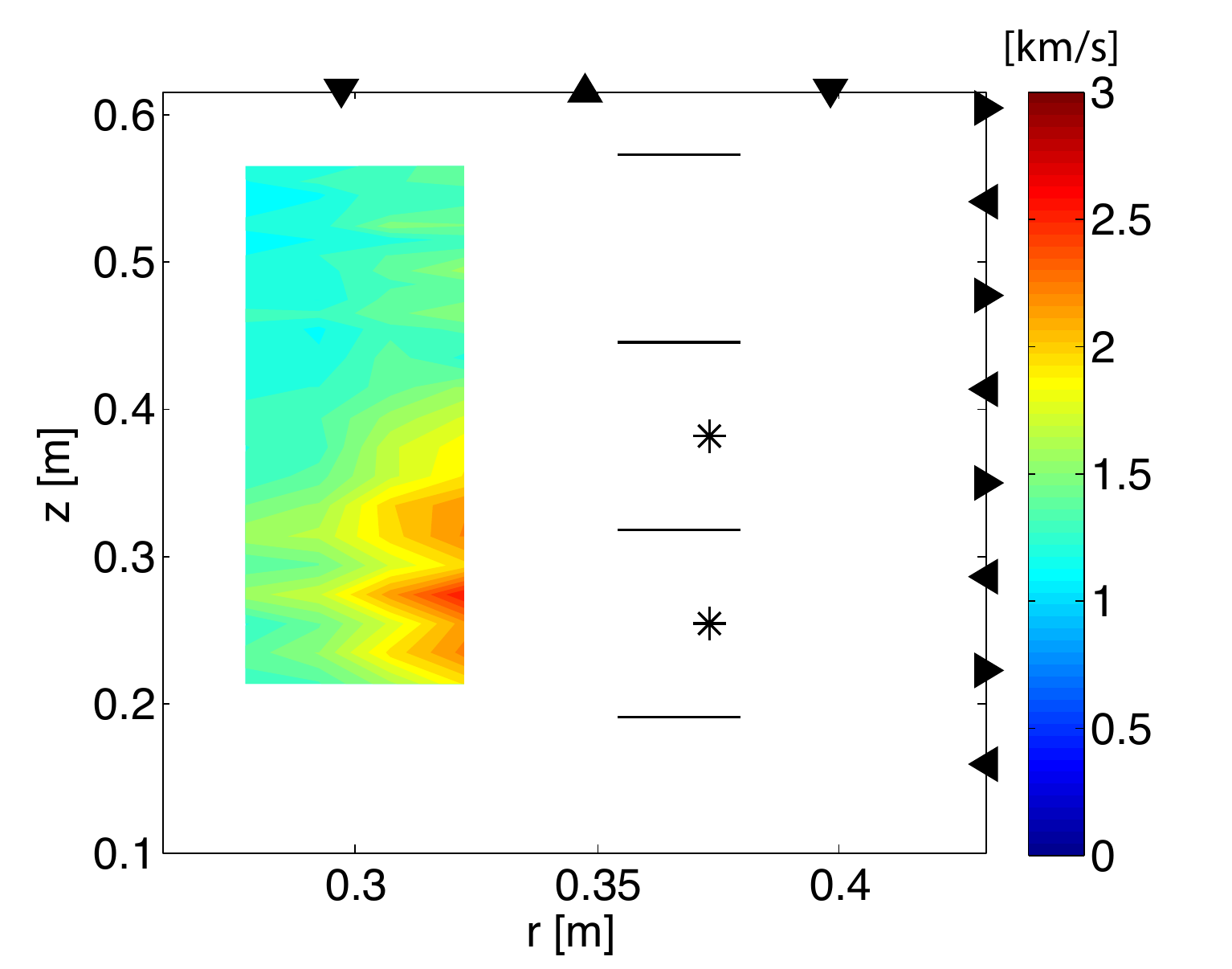}}
\subfloat[]{\label{fig:zfil2or4b}\includegraphics[width=\columnwidth]{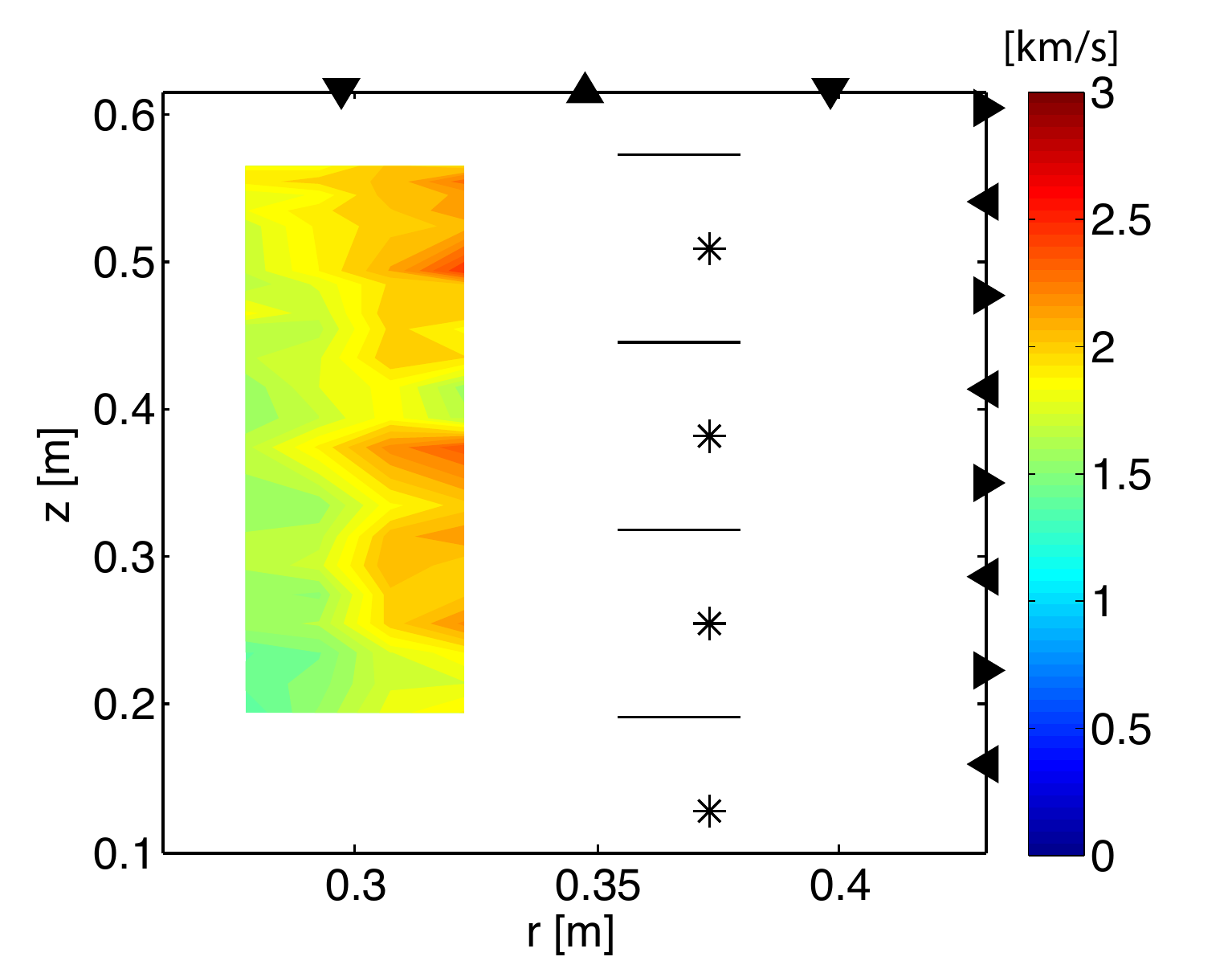}}
\end{center}
\caption{ Argon azimuthal velocity measured with vertically scanned, 4-tipped Mach probe, with 450 V for $\sim$ 1 A from each biased cathode with a) two biased cathodes vs. b) four biased cathodes. Triangles mark the location of magnets, anodes (solid lines) and cathodes (stars) are located around $r=0.37$ m.   }
\label{fig:zfil2or4}
\end{figure*}

The plasma is stirred using ${\bf J \times B}$ forces in the magnetized edge region, where current is driven by toroidally localized, electrostatically biased hot cathodes. Viscosity couples the azimuthal flow inward to the unmagnetized region, and the flow is optimized at low density (where viscosity is largest) and a low neutral pressure (where neutral drag is minimized). Mach probes measure flow using the ratio of upstream to downstream ion saturation current, $j^+/j^-=\mathrm{exp}(K V/c_s)$, with $K=1.34$. Measurements of the flow profiles match the profiles predicted by a model that includes Braginskii viscosity and momentum loss through ion-neutral charge exchange collisions \cite{collins2012_prl}. 

Rotation drive at the outer boundary has been extended to cover the entire vertical height of the plasma. In Fig.~\ref{fig:zfil2or4}, azimuthal velocity measurements were made with a vertically scanned, 4-tipped Mach probe in argon. The plasma parameters (measured by the triple probe) were similar for each case, with $n_e=3\times10^{10}$ cm$^-3$, $f_{ion}=10 \%$,  $T_e=6.5$ eV, and $T_i=0.1$ eV (from fitting the velocity profile). If all four outer cathodes are biased, the entire column of plasma spins. If only the two central cathodes are biased, rotation is induced exclusively at the midplane and exponentially decreases away from the midplane. Vertical shear is expected, since the kinematic viscosity of plasma is relatively high, $\nu \sim 3$ m$^2$/s, and $Re_{max}< 400$. For reference, the kinematic viscosity of inviscid fluids such as water or liquid metals \cite{morley2008} is $\nu \sim 10^{-6}-10^{-7}$ m$^2$/s. The plasma in PCX is more comparable to molasses, which has $\nu \sim 3-7$ m$^2$/s.

\section{Addition of Inner Boundary Flow Drive}
\begin{figure*}[t]
\begin{center}
\includegraphics[width=2.0\columnwidth]{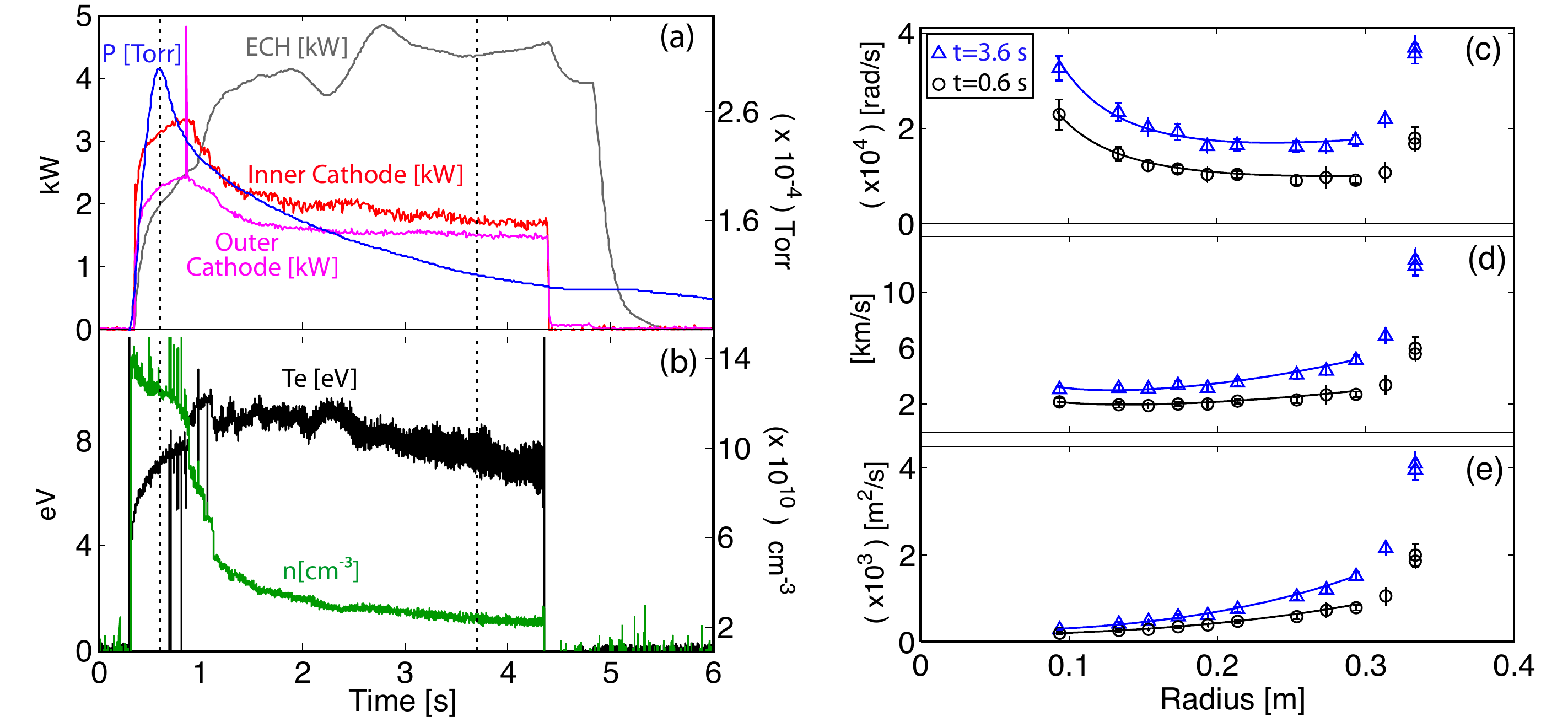}
\end{center}
\caption{ a) Plasma is created with a neutral gas puff and ECH heating. Rotation is induced with four outer cathodes (each 350 V, $\sim$ 1.1 A) and five inner cathodes (each 575 V, $\sim$ 0.75 A ). b) Plasma density and electron temperature vs. time, measured with the triple probe. c) Angular velocity profile at two different times during plasma discharge (times marked by dashed line in (a) and (b)), along with corresponding azimuthal velocity (d) and angular momentum (e) profiles. Error bars are standard deviation of fluctuation in time. }
\label{fig:he12kms} 
\end{figure*}

In order to control the shear flow necessary to excite flow-driven instabilities, rotation must also be induced at the inner boundary. Therefore, a center core assembly of magnets and stirring electrodes (pictured in Fig.~\ref{fig:PCXcollage}) was installed, consisting of 22 radially-magnetized (Br = 2.3 kG),  NeBFe magnet rings (OD=5.5 cm), stacked in alternating polarity and separated by aluminum clamps. Each inner cathode has a total surface area of $1.6 \mathrm{\ cm}^2$, constructed from loops of 0.038 cm diameter thoriated tungsten wire. 
Filaments are mounted by press-fit connections in double bore alumina tube with copper leads, and each cathode plugs in to a receptacle mounted on an aluminum clamp. Wiring was gauged to ensure that the filaments are the most resistive part of the circuit. Six out of the possible ten cathodes are installed; additional cathodes would require modification to both the center post and electrical vacuum feedthroughs.  The inner anodes are 2 cm long cylinders of $0.64$ cm diameter molybdenum, mounted on molybdenum threaded rod insulated by quartz tubes. All inner anodes are mounted to a copper bus bar and biased together. Cathodes are ohmically heated to temperatures around 1700$^{\circ}$C using variable-voltage AC power supplies and isolation transformers. The cathodes are negatively biased up to 600 V with respect to the cold anodes, and the entire electrode assembly is floating relative to ground (the chamber wall). The inner and outer electrode assemblies are biased independently with separate DC power supplies. 

The center post magnets were initially covered by a clear quartz tube (OD=6.02 cm). The aluminum center rod is water cooled to prevent magnets from overheating, but magnet surfaces still reached high temperatures of $\sim 120 ^\circ$C due to the radiant heat flux from filaments. Therefore, the inside of the clear quartz tube was coated with silver to act as a heat shield, resulting in acceptable magnet temperatures of $< 60^\circ$C during operation. Over time, the quartz tube becomes coated with a conductor, e.g. tungsten evaporating from hot filaments. In order to facilitate efficient cleaning and filament replacement, a gate valve system has been installed to isolate and remove the center core while preserving vacuum in the main chamber (see Fig. \ref{fig:PCXschematic}).  

In general, filaments most often fail in the overdense state of the plasma discharge ($n_e>7\times10^{10}$ cm$^{-3}$), during which large spikes in currents between biased cathodes and adjacent anodes are often observed (see Fig.~\ref{fig:he12kms}a , t=0.8 s). The inner cathodes were particularly vulnerable and tended to be destroyed by a single ``arc'' event. It is not clear if this was caused by shorting across the face of the magnets due to contamination by deposition of conducting material, or if it was a property of the cathodes themselves. One possibility is that higher plasma density could cause increased and possibly non-uniform heating through ion bombardment on the tungsten filaments, leading to localized hot spots and sudden bursts of emission current. Similar effects might occur because the coil shaped filaments could pick-up strong E-fields from microwaves in the edge region. 

Cathode behavior might be better understood through further studies of the edge region in overdense plasmas. Perhaps this problem can be solved by using a different cathode material and design, such as the LaB$_6$ cathode design currently being tested on the recently constructed, much larger Madison Plasma Dynamo Experiment \cite{Cooper_2014}. 

Basic characteristics of the microwave-heated, differentially rotating plasma are presented in Fig.~\ref{fig:he12kms}. Plasma is created using simultaneous gas puff and microwaves, and cathode bias is applied to both inner and outer cathodes. Density and electron temperature are measured with a triple probe. Azimuthal rotation profiles are measured with a radially scanned Mach probe at the midplane, and profiles are fitted with the Bessel function solution based on measured parameters\cite{collins2012_prl}. In the beginning of the shot, the plasma is overdense with $P=2.8 \times 10^{-4}$ Torr, $n_e=1.2 \times 10^{11}$ cm$^{-3}$, $T_e=6.8$ eV, and $T_i=1.1$ eV (found from the fitted velocity profile), resulting in $Pm=12$ and $Rm=30$ with peak flow speeds of $V_1=2.3$ km/s and $V_2=5.7$ km/s at the inner and outer boundaries, respectively.
Later in the shot, the plasma is underdense with $P=1.1 \times 10^{-4}$ Torr, $n_e=2.3 \times 10^{10}$ cm$^{-3}$, $T_e=7.5$ eV, and $T_i=0.3$ eV. At these parameters, peak flow speeds reached $V_1=3.4$ km/s and $V_2=12.3$ km/s, with $Pm=2.5$ and $Rm=65$. 

Maximum flow in the bulk was obtained when the neutral drag is minimized (at low neutral pressure and high ionization fraction) and ion viscosity $\nu$ is maximized (at low density, since $\nu \propto T_i^{5/2} n_i^{-1} \mu^{-1/2}$ for unmagnetized plasma). For the ${\bf J \times B}$ flow drive to be effective, the current has to be large enough and the electrode position in the magnetized edge must be optimized. The radial location for the center post stirring electrodes ($r \sim 7$ cm) was found through a series of trial-and-error experiments to test cathodes of various lengths.  

 \section{Measurements of Rotation-Induced Radial Electric Field}
\begin{figure*}[t]
  \begin{center}
    \includegraphics[width=.32\textwidth]{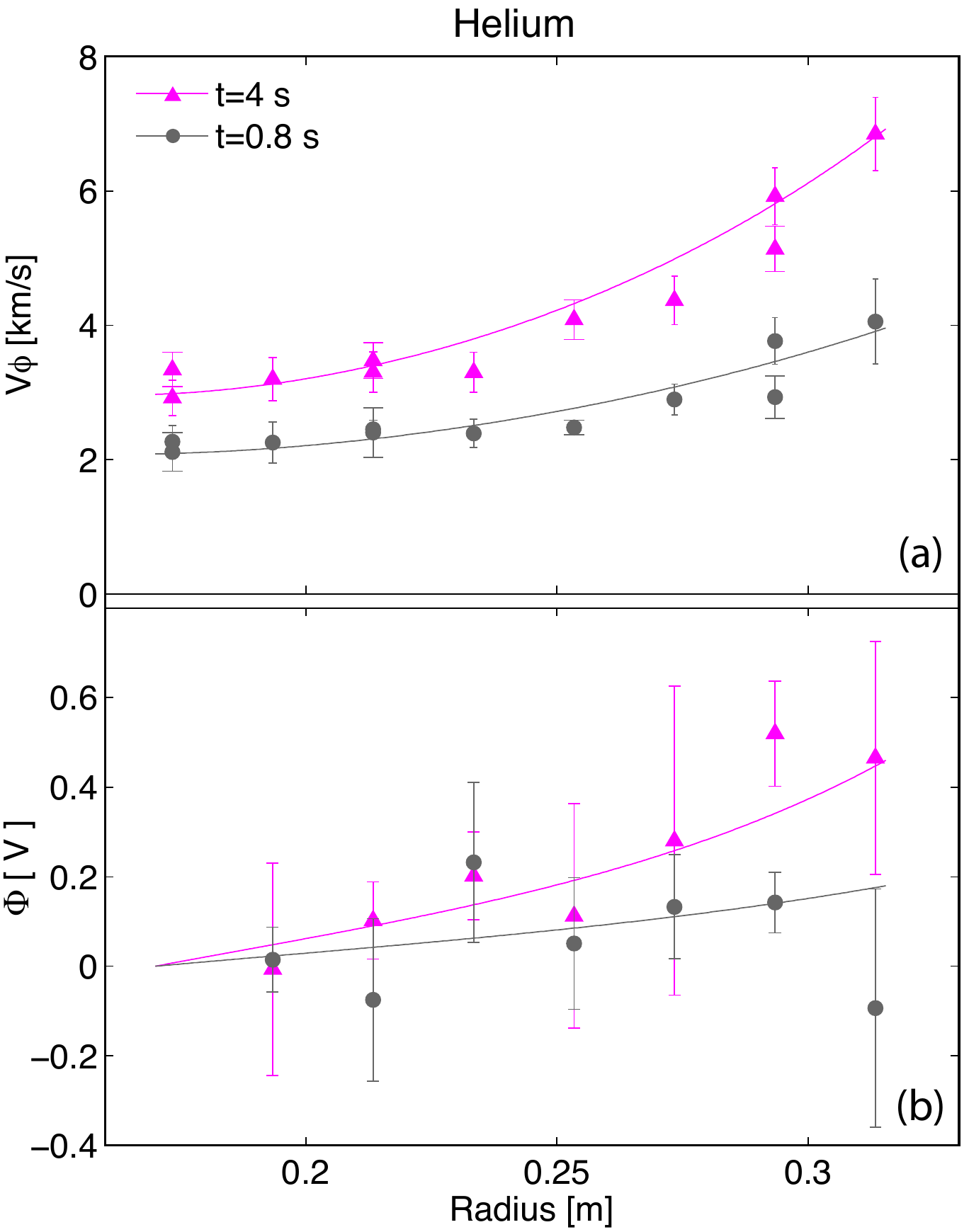}
    \includegraphics[width=.32\textwidth]{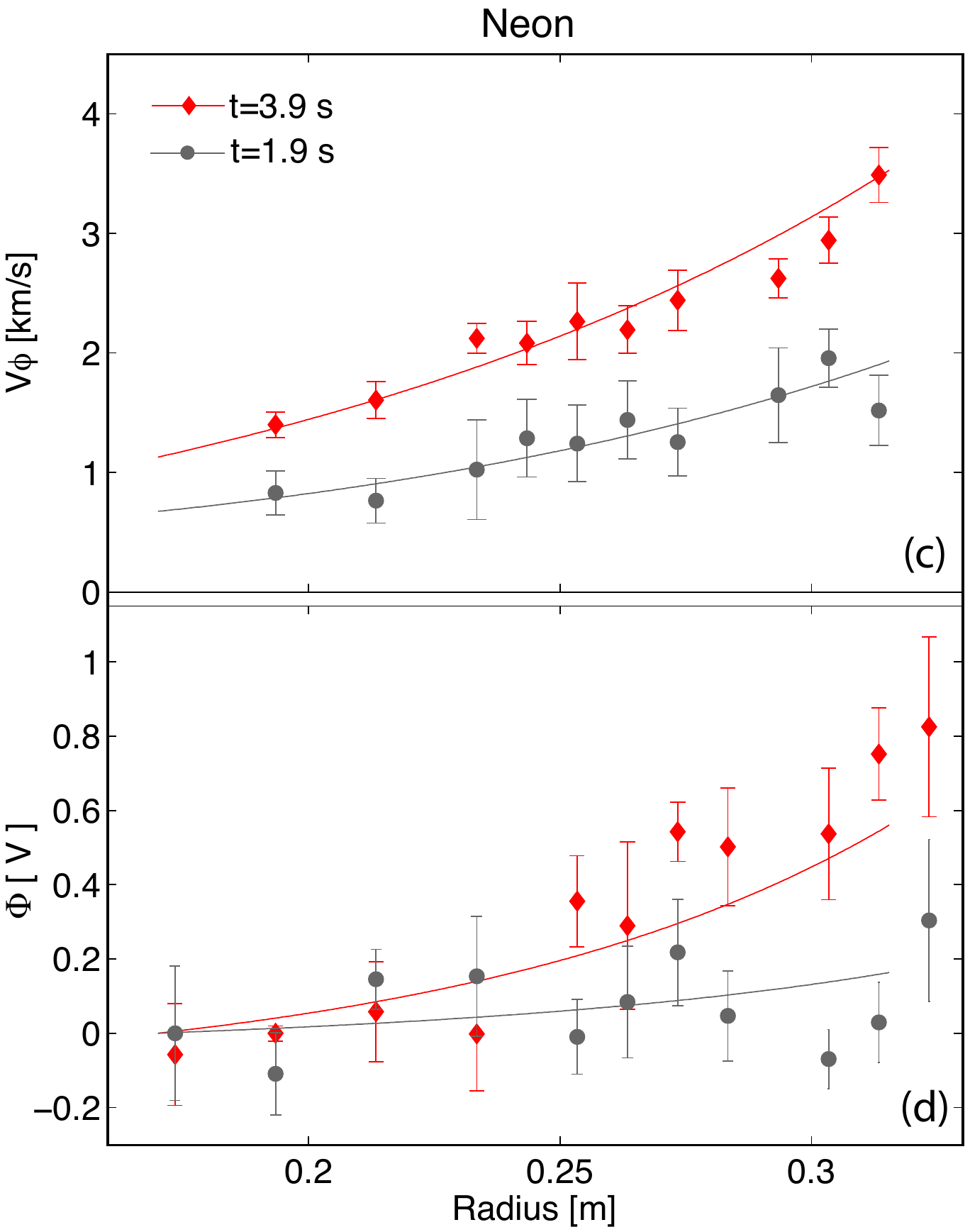}
    \includegraphics[width=.32\textwidth]{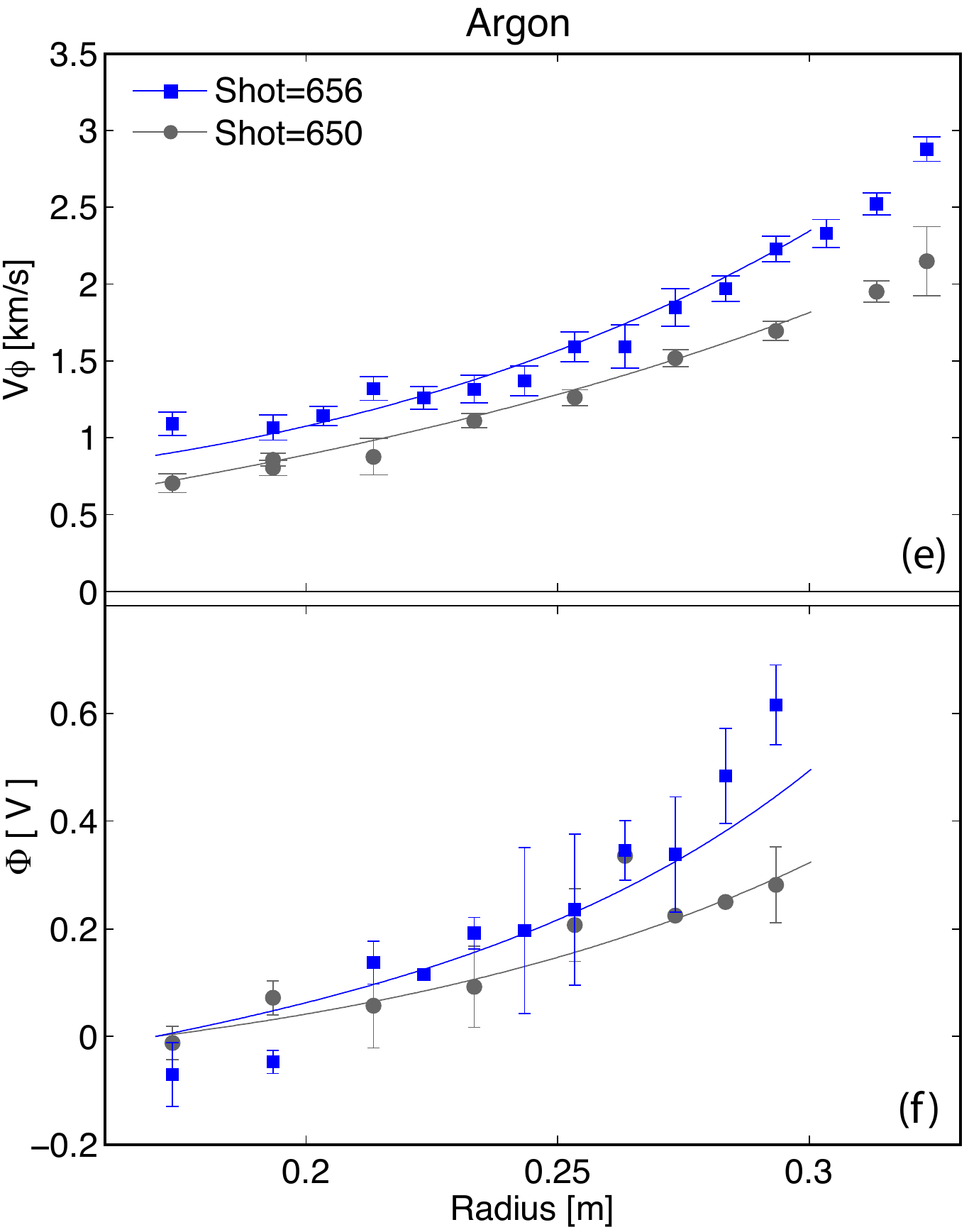}
\end{center}
\caption{Floating potential profiles (b,d,f) are measured for various flow profiles (a,c,e) generated in helium, neon, and argon. As the flow velocity increases, the radial floating potential increases as predicted by Eq.~\ref{eq:potential}. }
\label{fig:radialpotential} 
\end{figure*}

In a strongly rotating plasma, where the magnitude of the toroidal velocity of the plasma is comparable to the thermal velocity, the centrifugal force causes ions to move to larger radii. As with any gas centrifuge, the outward displacement of ions creates a pressure gradient, resulting in an inward force to balance the outward centrifugal acceleration. In a rotating plasma, however, a radial electric field arises to prevent charge separation from the difference in centrifugal forces \cite{wesson_1997}. The ions are thus held in orbit by both the ion pressure gradient force and the electric field as determined by the electron pressure. 

In the case of an unmagnetized flow with uniform temperature, the radial force balance equation for ions can be written:
\begin{equation}
- n_i M_i \frac{V_{\phi}(r)^2}{ r}  =  - Z e n_i \frac{d \Phi }{dr}  - T_i \frac{d n_i}{dr}
\label{eq:ioncentrifugal}
\end{equation}
where $ \Phi$ is the electrostatic potential. Neglecting the small centrifugal force on the electrons, the electrons simply obey the Boltzmann relation, $n_e = n_0 \mathrm{exp}( e \Phi / T_e)$. Since the electron density must match the ion density to maintain quasineutrality, a radial electric field arises. 
The resulting rotation-induced, mass-dependent plasma potential profile is
\begin{equation}
\Phi(r)  = \frac{1}{ e(Z + \frac{ T_i }{  T_e })  }  M_i \int_{R_1}^{R_2}   \frac{V_{\phi}(r)^2} { r}  dr 
 \label{eq:potential}
\end{equation}
where the inner and outer radii are denoted by $R_1$ and $R_2$. 

The PCX rotation drive scheme is unique in that it does not require external radial electric fields applied to the bulk, in contrast to most homopolar-type devices which create $E \times B$ flow. It would therefore be expected that a radial electric field appearing in the PCX plasma would be self-generated. The thermal diffusion gradient scale-length is much larger than the machine size and much smaller than the particle diffusion gradient scale-length in PCX, justifying the isothermal model assumed in Eq.~\ref{eq:potential}.  In the bulk, unmagnetized region away from any possible sources of electron heating (ECH resonance, joule heating due to biased cathodes), the electrons undergo multiple bounces, or reflections by the mirror force at the cusps, resulting in uniform distribution in the bulk. Uniform electron temperatures in the bulk, unmagnetized region (0.1 m $ < r < $ 0.3  m) in both rotating and non-rotating plasmas have been confirmed with both a radially scanned, swept Langmuir probe and OES, as shown in Fig. \ref{fig:shot690te}.

Potential profile measurements were obtained with a rake probe consisting of a single Mach probe along with multiple radially-spaced tips to measure floating potential. Assuming constant $T_e$,  $\Phi_{fl}$ and $\Phi_{p}$ vary by a constant sheath potential $\Phi_{sh}$, and $\Phi_{fl}$ can be substituted into Eq. \ref{eq:potential} by absorbing $\Phi_{sh}$ into the integration constant \cite{Hutchinson2002}. The rake probe measurements could be used to derive the radial electric field, however, the measurements tended to be very noisy ($| \Phi_{fl} | < 10$ V, with $\pm 2$ V fluctuations) and better results were obtained by simply measuring the potential profile and comparing to Eq.~\ref{eq:potential}.  Over the course of many plasma pulses, a number of measurements were recorded at each radial location, and the floating potential was then taken as the mean value between multiple shots. Efforts were made to maintain consistent plasma conditions, but the bulk plasma potential still varied slightly from shot-to-shot. To adjust for these shot-to-shot differences, the floating potential was simultaneously measured by the fixed triple probe and subtracted from the rake probe floating potential measurements. In order to extract the meaningful potential data (which is DC in character), a first order Savitzky-Golay smoothing routine (sgolayfilt in Matlab) has been applied to reduce the magnitude of the time-varying fluctuations and remove large voltage spikes from the data, which are often encountered in discharges with strongly biased cathodes.

Measurements of the radial floating potential show the presence of an outward rising potential which only arises when the plasma is flowing. The velocity and corresponding floating potential measurements in helium, neon, and argon are shown in Fig.~\ref{fig:radialpotential}.  The Mach probe velocity measurements are fitted with the Bessel function model based on the the measured plasma parameters for each dataset. The floating potential measurements are fitted with the predicted potential according to Eq.~\ref{eq:potential}, where the velocity profile is the Bessel function fit of the Mach probe velocity data. 

According to Fig.~\ref{fig:radialpotential}, the measured radial increase in potential is not much more than 0.8 V, and the corresponding density displacement is $10-20\%$ of the core density, according to $n_e/n_o=\mathrm{exp}(e\Phi/T_e)$. A density gradient has not yet been observed, because the interpretation of probe measurements based on collection of ion saturation current in a rotating plasma were ambiguous, and shot-to-shot variation in the measurement also tends to be on the order of $10-20\%$.

\subsection{Mass Dependence}
\begin{figure}[t]
\begin{center}
\includegraphics[width=.45\textwidth]{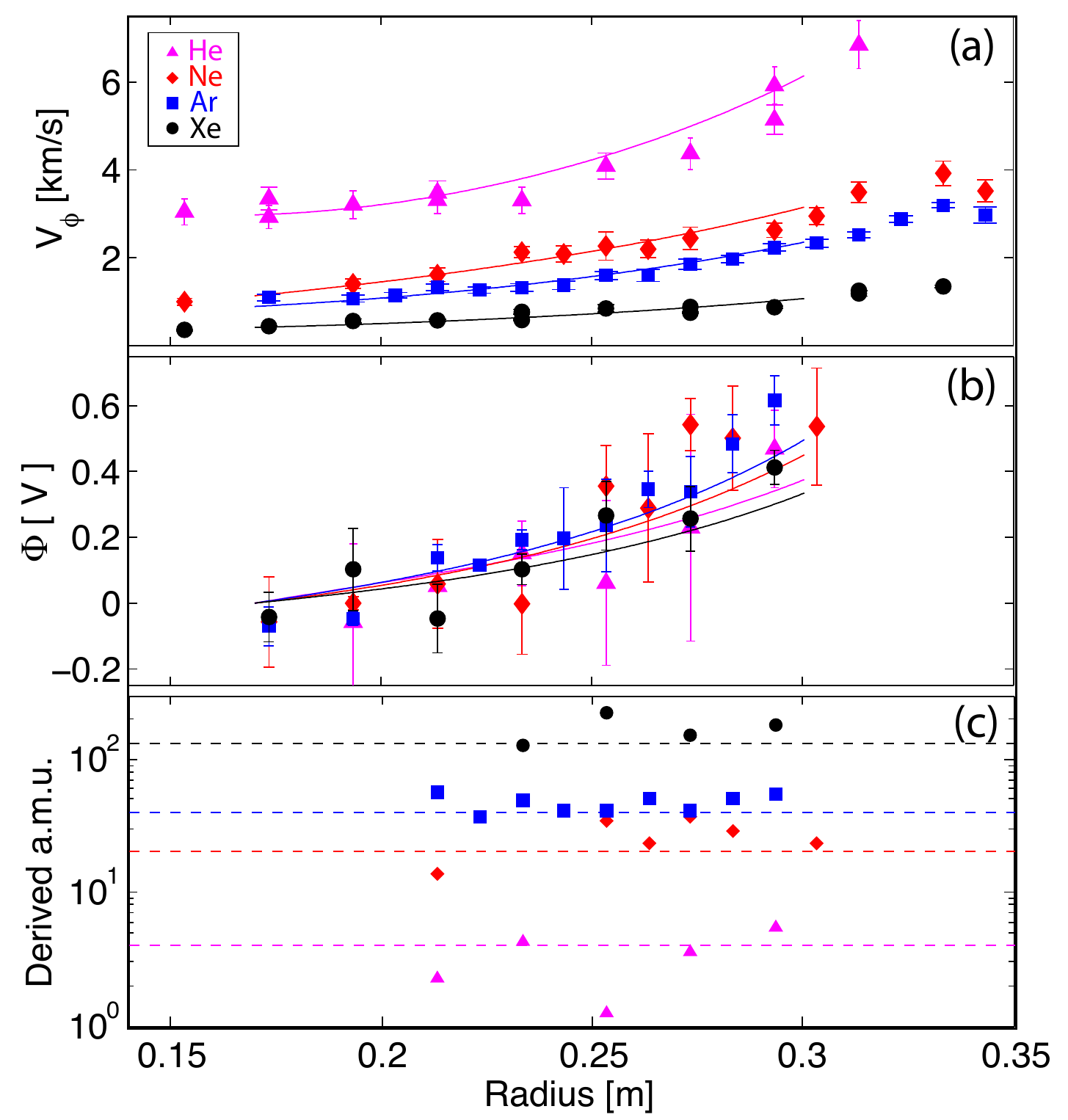}
\end{center}
\caption{The measured values of velocity and potentials in four different gases (helium, neon, argon, and xenon) result in derived ion masses (based on  Eq.~\ref{eq:ionmass} with Z=1)  which are close to the expected values (dashed lines). }
\label{fig:compareAll4} 
\end{figure}

The best evidence for mass-dependent, outward displacement of spinning plasma in PCX is found by comparing the profiles for flow and potential of various gases. One can solve for the ion mass using the measured potential and fitted velocity profiles:
\begin{equation}
M_i =\frac{e\Phi(r)(Z + \frac{ T_i }{  T_e })}{\int_{R_1}^{R_2}   \frac{V_{\phi}(r)^2} { r}  dr } 
 \label{eq:ionmass}
\end{equation}
In Fig.~\ref{fig:compareAll4}, experiments with four different gases (helium, neon, argon, and xenon) were
performed. Since the centrifugal force is stronger on heavier ions, lower rotation speeds are required to obtain radial potentials which are equivalent to the faster rotating, lighter species. The measured values of the potential and velocity are used to calculate the ion mass to within a factor of two for the distinct species.

\section{Velocity Limits}
The velocities measured in the PCX experiment have increased through the optimization of electrode locations and realization that plasma conditions are optimized at low density and high ionization fraction. It seems, however, that as long as the plasma remains partially ionized, velocity limits do exist and are consistent with the species-dependent critical ionization velocity first proposed by Alfv\'{e}n in 1954. Alfv\'{e}n hypothesized that if the ion flow kinetic energy exceeds the ionization potential of the neutrals, the ionization rate of a neutral gas should strongly increase. The critical ionization velocity (CIV) is reached when the relative velocity between the plasma ions and neutral gas reaches the value

\begin{equation}
V_{crit}=\sqrt{\frac{2eU_i}{m_n}}
 \label{civ}
\end{equation}
where $U_i$ is the ionization potential and $m_n$ is the mass of the neutral atom. Rather remarkably, the simple velocity limit set by Eq.~\ref{civ} has been confirmed in numerous experiments with a variety of geometries, spanning orders of magnitude in various neutral gas, plasma, and magnetic field conditions, as summarized in several extensive review articles \cite{brenning1992,lai2001}. Laboratory experiments that have been designed especially to study CIV generally create flowing plasma with either homopolar-type devices that implement cross-field discharges, or with coaxial plasma guns to create a plasma beam colliding with neutral gas. Investigations of many gas species found that the maximum $E/B$ velocity was typically within 50$\%$ of $V_{crit}$ (as long as the plasma remained partially ionized), where the magnitude of the velocity was inferred by $|E/B|$ or by the Doppler shift of spectral lines.  

The peak velocities measured in PCX for four different gas species (helium, neon, argon, xenon) are listed in Table \ref{tab:peakvelocitytable}. The velocities were measured with a Mach probe at the midplane near the outer edge of the unmagnetized region, around $r=0.34$ m. Density and electron temperature were measured with both the single swept Langmuir probe and triple probe. In all cases, the peak speeds were always less than the calculated $V_{crit}$ for each species. While $V_{max} < V_{crit}$, it is conceivable that peak velocities reached higher speeds in regions of the plasma that were not measured by the Mach probe, for example directly in front of the permanent magnets where $ \bf{J\times B}$ forces are large. Attempts were made to further increase the velocity by increasing $\bf {J}$ (by increasing the filament temperature and bias). This resulted in increased density and often caused the plasma to switch from the underdense to overdense state in microwave discharges, leading to undesirable filament arcing. Interestingly, the ratio of $V_{max}/ V_{crit}\sim 0.3$ for all gases tested. Thus, it seems that a velocity limit which depends on the species mass and ionization potential, consistent with CIV phenomenon, is observed in PCX. 

\begin{table}
\centering
\resizebox{.45\textwidth}{!}{
\footnotesize
\begin{tabular}{lccccccccc}
	\hline
	\\[1pt]
$$\bf{Gas}  & $\bf{U_i}$ &$\bf{M_i}$& $\bf{V_{crit}}$ & $\bf{V_{max}}$ &$\bf{ n_e \times 10^{10}}$  & $\bf{f_{ion}}$ &  \bf{Te}  &$\bf{Ti_{fit}}$ \\
       & (eV)   &(a.m.u)& (km/s)  &(km/s)          &(cm$^{-3}$)          &($\%$)       &(eV)   & (eV) \\            	
	\hline
He    & 24.6 & 4         &34	  & 12               &  2.3     & 0.7      &6.9   &0.21   \\
Ne    & 21.6 & 20.2    &14	  & 4                   &  3.4     & 1.2      &7.7   &0.3      \\
Ar     & 15.8 & 40       &8.7	  & 3.2                 &  3        & 9.5       &5.9  &0.135  \\
Xe    & 12.1 & 131.3   &4.2	  & 1.35                & 6         & 32        &3.6  &0.15    \\
        \hline
\end{tabular}}
\caption[Peak Velocities in PCX]{Summary of ionization energy, mass, critical ionization velocity, maximum induced velocities for different gas species in PCX as measured by the Mach probe. Also listed for reference are the corresponding $n_e$, $T_e$, and the ionization fraction $f_{ion} \equiv n/n_o$.  Note that $V_{max}$ is always less than $V_{crit}$, with $V_{max}/ V_{crit}\sim 0.3$ in all cases. }
\label{tab:peakvelocitytable}
\end{table}

\section{Hydrodynamic Stability of Taylor-Couette Flow of Plasma}
In PCX, flow which is driven exclusively at the outer boundary results in nearly solid-body rotation, resembling a centrifuge. Such profiles are hydrodynamically stable because the angular momentum increases with radius , and the outward centrifugal force is balanced by the inward pressure gradient force in accordance with the Rayleigh stability criterion. As sheared flow is driven by inner boundary flow drive, the flow can become hydrodynamically unstable above a certain critical threshold. The fluid stability problem of TCF can be formulated using traditional linear stability methods \cite{Chandrasekhar_1961}, except with the inclusion of the plasma-specific ion-neutral drag term. In the unmagnetized bulk region, the toroidal force balance is given by 
\begin{eqnarray}
\frac{\partial \mathbf{V}}{\partial t} + (\mathbf{V} \cdot \nabla) \mathbf{V} = -\frac{\nabla P}{\rho} -\frac{\mathbf{V}}{\tau_{i0}} + \nu \nabla^2 \mathbf{V}  \qquad
\end{eqnarray} 
where $\nu$ is viscosity due to ion-ion collisions and $\tau_{i0}  \approx (n_0 \langle\sigma_{cx} v\rangle)^{-1} $ is the ion-neutral collision time; $\sigma_{cx}$ is the charge-exchange cross-section. Assuming axisymmetry with a perturbed flow $\mathbf{V+ u } = [ u_r, V_0+u_\theta , u_z]$ and $\bar{\omega}=\frac{\delta P}{\rho}$ , the linearized momentum equation is:
\footnotesize
\begin{eqnarray}
&&\frac{\partial u_r}{\partial t}-2\frac{V_0 u_{\theta}}{r}=  -\frac{\partial \bar{\omega}}{\partial r} - \frac{u_r}{\tau_{i0}}+ \nu\left(\nabla^2 u_r - \frac{u_r}{r^2}\right)\qquad \\
&&\frac{\partial u_{\theta}}{\partial t}+\left ( \frac{V_0}{r}+ \frac{\partial V_0}{\partial r} \right ) u_r = -\frac{u_{\theta}}{\tau_{in}}+\nu\left(\nabla^2 u_{\theta}-\frac{u_{\theta}}{r^2}\right)\quad \\
 &&\frac{\partial u_z}{\partial t}=-\frac{\partial \bar{\omega}}{\partial z}-\frac{u_z}{\tau_{in}}+\nu \nabla^2 u_z 
\label{eq:linmom}
 \end{eqnarray} 
 \normalsize
As shown previously, TCF of partially ionized plasmas is modified by drag induced from plasma ion-neutral charge exchange collisions. Hence, the base flow $V_0$ is the equilibrium ion-neutral damped plasma TCF profile given by modified Bessel functions \cite{collins2012_prl} 
\begin{equation}
V_0(r)=AI_1(r/L_v)+BK_1(r/L_v),
\end{equation}
where the constants $A$ and $B$ are determined by the boundary values of the velocity and $L_v=\sqrt{\tau_{i0}\nu}$. Suppose perturbations of the form
 \begin{align}
 u_r &=e^{\gamma t} u(r) \mathrm{cos}(kz)\\
 u_\theta &= e^{\gamma t} v(r) \mathrm{cos}(kz)\\
 u_z &= e^{\gamma t} w(r) \mathrm{sin}(kz)\\
 \bar{\omega}&= e^{\gamma t}\bar{\omega}(r)\mathrm{cos}(kz)
 \end{align} 
where the growth rate is $\gamma$, the system is periodic in the z-direction (described by the axial wavenumber $k$), and the perturbation amplitudes depend on the radial position.  Assuming incompressible flow and using the notation $\frac{d}{dr} \rightarrow D$ and $\frac{d}{dr}+\frac{1}{r} \rightarrow D_*$, we find 
 \begin{align}
 \frac{\nu}{k^2}\left[DD_*-\left(k^2+\frac{1}{L_v^2}+\frac{\gamma}{\nu}\right)\right] \times & \nonumber \\ 
 \left(DD_*-k^2\right)u  =& \frac{2 V_0}{r}v& \\
\nu\left[DD_* - \left(k^2+\frac{1}{L_v^2}+\frac{\gamma}{\nu}\right)\right]v =& (D_*V_0) u & 
 \end{align}

\begin{figure}
\begin{center}
\includegraphics[width=.45\textwidth]{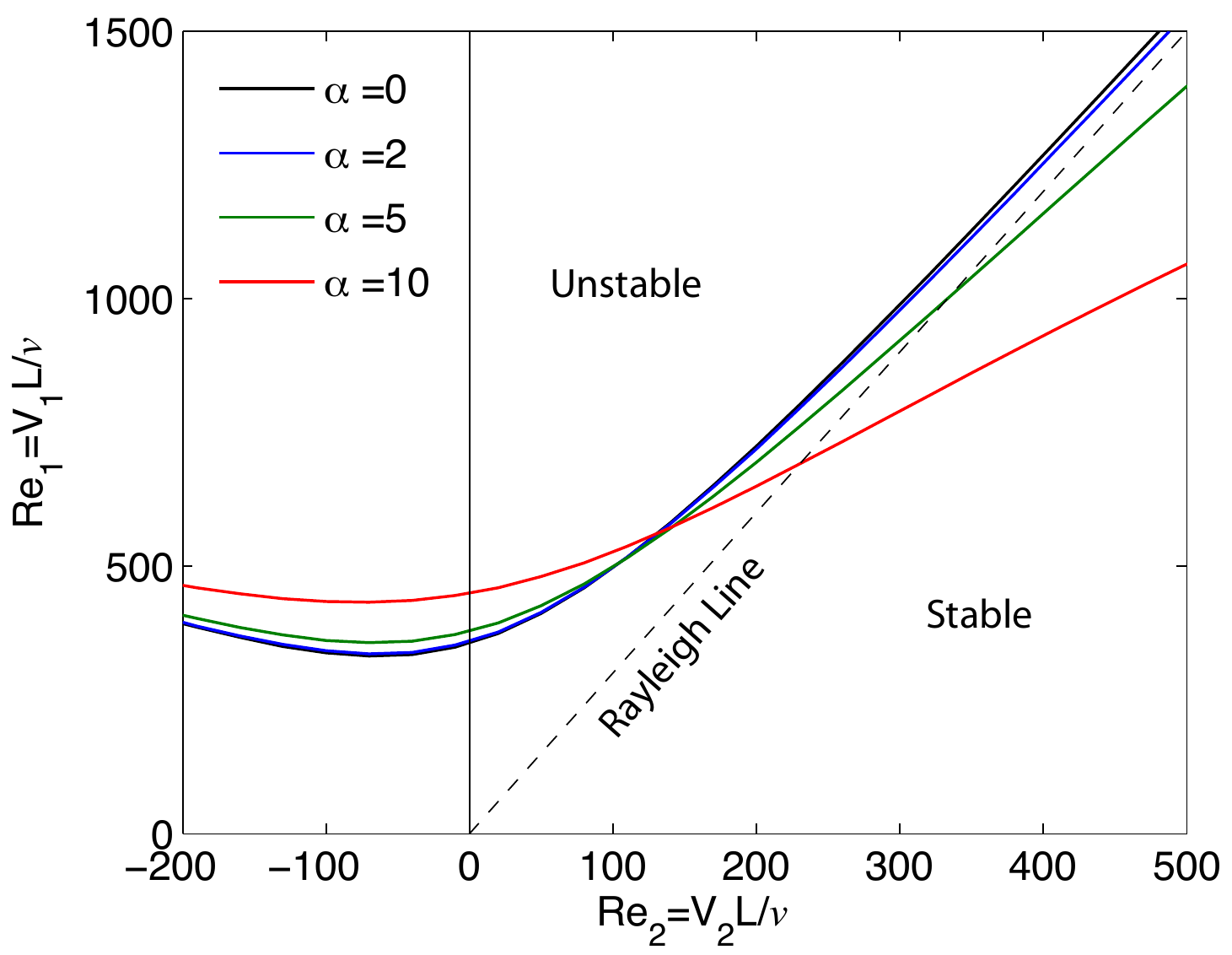}
\end{center}
\caption{Marginal stability diagram for Taylor-Couette flow. As the momentum diffusion length decreases, flow becomes unstable in regions which would otherwise be Rayleigh stable. }
\label{fig:TCstabilityVS3alpha} 
\end{figure}

\begin{figure}
\begin{center}
\includegraphics[width=.45\textwidth]{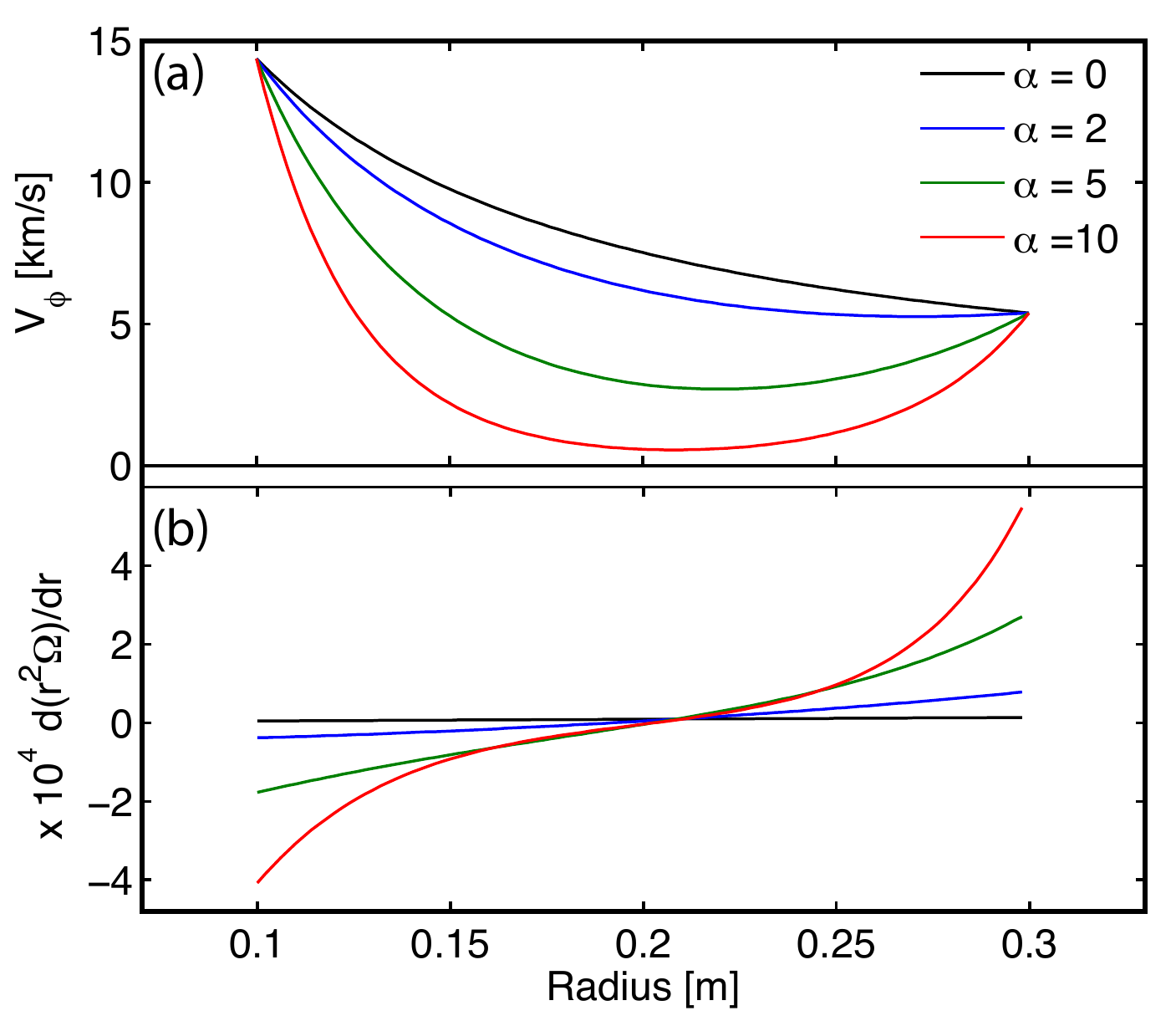}
\end{center}
\caption{Top) Representative flow profiles for various momentum diffusion lengths with $Re_1=800$ and $Re_2=300$.  Bottom) Corresponding calculation of $d(r^2\Omega)/dr$. Flows can be unstable in regions where perturbations are not damped by viscosity and $d(r^2\Omega)/dr<0$.}
\label{fig:Vphi_dL_3alpha} 
\end{figure}

These equations can then be normalized, discretized, transformed into an eigenvalue problem at marginal stability (where $\gamma=0$), and solved numerically. The numerical calculation is in terms of the normalized momentum diffusion length parameter, $\alpha=L/L_v$, or 
\begin{align}
\resizebox{.85\columnwidth}{!}{$\alpha  =  153 \; L_ \text{m} Z^2 \sqrt{\sigma_{\text{cx},10^{-19} \text{m}^2} n_{ \text{i,}10^{11} \text{cm}^{-3}}}\;T_{\mathrm{i,eV}}^{-1} $}
\end{align}
where $\sigma_{\mathrm{cx},10^{-19}\mathrm{m}^2}$ is the charge exchange cross-section in units of $10^{-19}$m$^2$. The assumed axial wavenumber is $k=\pi/(R_2-R_1)$ with PCX dimensions $R_1=0.1$ m and $R_2=0.3$ m. The eigenvalue solver then finds the critical Reynold's number at the inner boundary ($Re_1=V_{1}L/\nu$) for each prescribed $Re_2$.

The results of the marginal stability analysis are shown in Fig.~\ref{fig:TCstabilityVS3alpha}. Stability curves for several values of $\alpha$ are plotted in the parameter space of dimensionless fluid Reynold's numbers of the inner and outer cylinders ($Re_1$, $Re_2$). At small values of $\alpha$, ion-neutral collisions are unimportant, and the azimuthal velocity profile takes on the classical form for TCF of a viscous fluid, where $\alpha = 0$. In this limit, the minimum critical $Re_1\sim 330$ is determined by viscosity, and the stability curve asymptotes to Rayleigh's prediction for stability of inviscid fluid ($\Omega_1R_1^2=\Omega_2 R_2^2$), which is plotted as a dashed line in Fig.~\ref{fig:TCstabilityVS3alpha}. 

As $\alpha$ increases, ion-neutral collisions become more important, and the plasma flow becomes unstable in regions which would otherwise be Rayleigh stable in classical fluid flows. This is due to the increased shear in the velocity profile created by momentum loss in ion-neutral collisions. In Fig.~\ref{fig:Vphi_dL_3alpha}, profiles of flow and $d(r^2\Omega)/dr$ are plotted for various neutral gas pressures ($8\times 10^{-6} - 2 \times 10^{-4}$ Torr), which correspond to different values of $\alpha$. In this example, $Re_1=800$ and $Re_2=300$, and azimuthal velocity profiles are calculated based on fixed helium parameters $L=0.3$ m, $n=4 \times 10^{10}$ cm$^{-3}$, $T_e=8$ eV, and $T_i=0.1$ eV.  Centrifugal instability can occur in regions where $d(r^2\Omega)/dr<0$. Note that high neutral pressure can also increase the instability threshold for rotation driven exclusively at the inner boundary. Thus far, shear flows in PCX have been generated at high helium neutral pressures ($>10^4$) and lower inner boundary velocity, where $Re_1<10$. According to the marginal stability analysis, the achieved flows would be expected to be hydrodynamically stable. 

\section{Summary}
The demonstration of sheared flow of plasma in the unmagnetized core region, driven by biased electrodes in the magnetized plasma edge, is a major advance for carrying out future flow-driven MHD experiments. A number of experimental observations have been reported here:  
\begin{itemize}
\item A self-consistent, rotation-induced radial electric field is measured. It only appears when there is flow, it scales with the magnitude of the flow and the ion mass, and it is in agreement with the predicted profile calculated from fitted velocity measurements.  
\item The $T_e$ measurements of Langmuir probes have been independently confirmed by a newly developed Optical Emission Spectroscopy technique, leading to further confidence in the \emph{magnitude} of the Mach probe velocity measurements.  Furthermore, the $T_e$ profiles are flat, so the changes in potential profiles are an important confirmation of Mach probe measurements.
\item Maximum velocity limits have been observed for various gas species ( He $\sim$ 12 km/s, Ne $\sim$  4 km/s, Ar $\sim$ 3.2 km/s, Xe $\sim$ 1.4 km/s ), consistent with the well documented critical ionization velocity limit phenomenon, which exists in partially ionized plasma.
 \item Flow can also be driven at the inner boundary, where the geometry is slightly different with smaller cathodes and stronger center stack magnets.
\end{itemize}

The next step for the experiment is to expand the parameter space by further maximizing flow speeds at the inner boundary and driving flow at higher densities. Rotation would improve at increased ionization fractions, which could be accessed through additional vacuum pumping and improving confinement, for example by  replacing the first generation of ceramic magnets on the endcaps and outer boundaries with a second generation of the high-field, neodymium type. Spinning at higher density may also require improvements to cathode design. Additional microwave power and launching of X-mode waves ($\mathbf{E}\perp\mathbf{B}$), which are not restricted by density cutoffs, could also be explored.

\section{Acknowledgments}

This work was funded in part by NSF award AST-1211937 and the Center for Magnetic Self Organization in Laboratory and Astrophysical Plasmas. C.C. acknowledges support by the ORISE Fusion Energy Sciences Graduate Fellowship.

\end{document}